\documentclass[twocolumn,prd,aps,showpacs,amsmath,amssymb]{revtex4-1}
\usepackage{graphics}
\usepackage{graphicx}
\usepackage{dcolumn}
\usepackage{bm}
\usepackage{mathrsfs}
\usepackage{pstricks}
\usepackage{color}
\usepackage{slashed}
\usepackage{amsmath}
\usepackage{epsfig}
\usepackage{amsfonts}
\usepackage{amssymb}

\def\sgn{\mathrm{sgn}}
\def\beq{\begin{equation}}
\def\eeq{\end{equation}}
\def\bea{\begin{eqnarray}}
\def\eea{\end{eqnarray}}
\def\nn{\nonumber}
\begin{document}

\title{Radiative processes of uniformly accelerated entangled atoms}
\author{G. Menezes}
\email{gabrielmenezes@ufrrj.br}
\affiliation{Grupo de F\'isica Te\'orica e Matem\'atica F\'isica, Departamento de F\'isica, Universidade Federal Rural do Rio de Janeiro, 23897-000 Serop\'edica, RJ, Brazil}
\author{N. F. Svaiter}
\email{nfuxsvai@cbpf.br}
\affiliation{Centro Brasileiro de Pesquisas F\'{\i}sicas, 22290-180 Rio de Janeiro, RJ, Brazil}

\begin{abstract}
We study radiative processes of uniformly accelerated entangled atoms, interacting with an electromagnetic field prepared in the Minkowski vacuum state. We discuss the structure of the rate of variation of the atomic energy for two atoms travelling in different hyperbolic world lines. We identify the contributions of vacuum fluctuations and radiation reaction to the generation of entanglement as well as to the decay of entangled states. Our results resemble the situation in which two inertial atoms are coupled individually to two spatially separated cavities at different temperatures. In addition, for equal accelerations we obtain that one of the maximally entangled antisymmetric Bell state is a decoherence-free state.

\end{abstract}


\maketitle

\section{Introduction}
\label{intro}

Superposition and entanglement are the distinguished properties of the quantum theory. Entangled pure quantum states are defined as states that cannot be factorized into product of pure states of subsystems. Entanglement has attracted much interest since it is a essential feature underlying quantum information, cryptography and quantum computation~\cite{1,haroche}. There exist several sources of entangled quantum systems, such as in solid state physics, quantum optics and also atoms in cavity electrodynamics. Many ways were proposed in order to generate entangled states in systems of two-level atoms interacting with a bosonic field. See for example~\cite{2,3,4,5,ved}.

In a seminal paper, Yu and Eberly studied the radiative processes of entangled two-level atoms coupled individually to two spatially separated cavities~\cite{eberly}. A different scenario was discussed in Ref.~\cite{juan}. In this work, radiative processes of a quantum system composed by two identical two-level atoms interacting with a massless scalar field prepared in the vacuum state in the presence of perfect reflecting flat boundaries were analyzed. The authors investigated the spontaneous transitions rates from the entangled states to its collective ground state induced by vacuum fluctuations. In the presence of a single boundary the transition rate for the symmetric state undergoes a strong reduction, whereas for the antisymmetric state the results indicate a slightly enhancement. Following the discussion by Ackerhalt, Knight and Eberly~\cite{ake}, which presents the possibility to interpret spontaneous decay as a radiation-reaction effect, in Ref.~\cite{ng1} we have discussed quantum entanglement between inertial atoms within the formalism developed by Dalibard, Dupont-Roc, and Cohen-Tannoudji (DDC)~\cite{cohen2,cohen3}. As discussed by Milonni the effects of vacuum fluctuations and radiation reaction depend on a particular ordering chosen for commuting atomic and field operators~\cite{mil1}. The DDC formalism asserts that there is a unique way in which such distinct contributions to the rate of change of an atomic observable are separately Hermitian and hence they enjoy an independent physical meaning. This is achieved by
choosing a preferred operator ordering, namely a symmetric ordering of commuting atom and field variables. Within such a formalism the interplay between vacuum fluctuations and radiation reaction can be considered to maintain the stability of the atom in its ground state. We remark that this formalism was employed in many situations~\cite{aud1,aud2,passante,china1,china2,rizz}. For uniformly accelerated atoms, we remark that this method motivates in a quantitative sense the intuitive scenario outlined in Ref.~\cite{sciama}.

Recently there has been an extensive literature devoted to investigations of quantum information theory within the framework of quantum field theory and quantum field theory in curved space-time. This emerging field of relativistic quantum information has been attracting increasing interest within the scientific community. In its essence, this subject aims at understanding the role that relativistic settings play in quantum information processes. In some way, familiar notions in quantum information theory claim for a critique reappraisal when one brings up relativistic considerations. Even though we do not intend to give an extensive bibliography of the subject, here we call attention on some of such important works. For instance, in reference~\cite{fue} it was shown that entangled states reduce to separable states at infinite acceleration. This phenomenon is sometimes called Unruh decoherence and is closely related to the Unruh effect~\cite{birrel,nami,mat}. The idea is that the acceleration of the observer causes a sort of enviromental decoherence which narrows the fidelity of certain processes of primary importance in quantum information. Relationships between entanglement and space-time curvature were investigated in Ref.~\cite{men}. The degradation phenomenon of entanglement in a Schwarzschild black hole was investigated in Ref.~\cite{mar}. On the other hand, in contrast to this effect, in references~\cite{mira1,mira2} the authors considered a situation of tripartite entanglement in which one of the three entities moves with a uniform acceleration with respect to the others; their results indicate that entanglement does not completely vanish in the infinite acceleration limit. In turn, generation of quantum entanglement between two localized causally disconnected atoms was studied in references~\cite{rez1,rez2,braun,mass,fran,LinHu2009,LinHu2010}. Similar results were found in references~\cite{ben1,ben2,china3,china4}, but with different setups and within distinct scenarios. We also mention investigations of quantum teleportation between noninertial observers~\cite{als1,als2,shiw}, relativistic approaches to the Einstein-Podolsky-Rosen (EPR) construction and also to Bell's inequality~\cite{cza,terno,tera,kim,cab}, EPR-type correlations in quantum fields~\cite{summers} and entanglement in fermionic fields~\cite{als3,ber}. See also~\cite{peres,ging,ging2,shi,cza2,jor,ade,ani} for more interesting discussions on the subject of relativistic quantum entanglement. As a general source for relativistic quantum information we also refer the reader the Ref.~\cite{peres2} which also indicates some open problems in such a research field. Various  aspects of this important research topic are reviewed in the special issue of {\it Classical and Quantum Gravity} [{\bf 29} (22) (2012)]. One of the most important implications of such studies is the realization of the observer-dependency property of quantum entanglement~\cite{als4}. Hence a quantitative understanding of such phenomena is highly required in discussions on certain quantum information processes between accelerated members of a system.

In the description of the dynamics of a system coupled to a reservoir different formalisms have been used. Generally the reservoir is described by an infinitely many degrees-of-freedom formalism. The standard framework is the traditional master equation approach. The evolution of the reduced density matrix of the atoms is usually written in the Kossakowski-Lindblad form. Within this approach it is possible to quantify the degree of entanglement of a particular state. Here since we are interested in apprehending the machinery responsible for generation (or degradation) of entanglement in radiative processes involving atoms, we choose a different route and use the DDC formalism aforementioned. Moreover, this formalism easily enables one to discuss classical and quantum mechanical concepts. Hence in the present paper we propose to generalize the results of Ref.~\cite{china5} in which the authors investigate quantum entanglement between two uniformly accelerated atoms with the same acceleration. Being more specific, we assume that both atoms travel in different hyperbolic trajectories. We consider that both atoms are coupled to a quantum electromagnetic field. In addition, we intend to identify the role of vacuum fluctuations and radiation reaction in the radiative processes of entangled atoms in uniformly accelerated motion.

The organization of the paper is as follows. In Section~\ref{model} we discuss the implementation of the DDC formalism in the situation of interest. In Section~\ref{atom} we calculate in detail the distinct contributions of vacuum fluctuations and radiation reaction to the rate of variation of atomic energy in vacuum for atoms moving along different hyperbolic trajectories. Specific discussion regarding the obtained results is given. Conclusions and final remarks are given in Section~\ref{conclude}. The
paper includes Appendices containing details of lengthy derivations. In this paper we use units $ \hbar = c = k_B = 1$. We are using the Minkowski metric $\eta_{\alpha\beta} = -1, \alpha=\beta=1,2,3$, $\eta_{\alpha\beta} = 1, \alpha=\beta=0$ and $\eta_{\alpha\beta} = 0,\alpha \neq \beta$. All graphics were drawn using the Mathematica package.

\section{Two identical atoms coupled with a electromagnetic field}
\label{model}

Let us suppose the case of two identical two-level atoms interacting with a common electromagnetic field. We are working in a four-dimensional Minkowski space-time. In this paper we work in the multipolar coupling scheme which means that all interactions are realized through the quantum electromagnetic fields. This formalism is suitable for describing retarded dipole-dipole interactions between the atoms. Let us consider that such atoms are moving along different hyperbolic trajectories in the $(t,x_1)$ plane. In this case, the atoms are said to be Rindler observers. As well known, the Rindler metric $ds^2 = e^{2a\xi}(d\eta^2 - d\xi^2) - d{\bf x}_{\perp}^2$ is the metric seen by an uniformly accelerated observer. Rindler coordinates $(\eta,\xi)$ and Minkowski coordinates $(t,x_1)$ are related by the known relations~\cite{birrel}
\bea
&& x_1 = a^{-1}e^{a\xi}\cosh a\eta
\nn\\
&& t = a^{-1}e^{a\xi}\sinh a\eta,
\label{acc}
\eea
where $a$ is a positive constant and $-\infty < \eta,\xi < \infty$. The coordinates $(\eta,\xi)$ cover only a quadrant of Minkowski space, namely the wedge $x_1 > |t|$. Lines of constant $\eta$ are straight, whereas lines of constant $\xi$ are hyperbolae:
$$
x_1^2 - t^2 = a^{-2}e^{2a\xi}.
$$
These represent world lines of uniformly accelerated observers. One has that the (constant) proper acceleration is given by
\beq
a e^{-a\xi} = \alpha.
\eeq
Thus, different (but constant) $\xi$'s correspond to different hyperbolae and hence to different proper accelerations. The accelerated observers' proper time $\tau$ is related to $\xi,\eta$ by
\beq
\tau = e^{a\xi}\eta.
\label{propt}
\eeq
This implies that in general, the atoms will be moving along different world lines, so there will be two different proper times parameterizing each of these curves. We remark that the quantity $a$ remains the same for both atoms in this situation.

Here we would like to identify the distinct contributions of quantum field vacuum fluctuations and radiation reaction to the entanglement dynamics of the atoms. As discussed above, we consider both atoms on different stationary trajectories $x^{\mu}(\tau_i) = (t(\tau_i),{\bf x}(\tau_i))$, where again $\tau_i$ denotes the proper time of the atom $i$. Usually one employs the proper time in order to describe time evolution of the atom-field system since it is directly measurable by the clocks of the observers. Here, because the atoms are moving along distinct hyperbolic trajectories, one has two proper times. However, to derive the Heisenberg equations of motion of the coupled system, we have to choose a common time variable. This implies that a certain amount of care is required when choosing the parameter which will describe the time evolution of the system. At first sight, it is most reasonable to refer generally to one of the atom's proper time $\tau_1$ and then relate the other proper time $\tau_2$ with $\tau_1$. At any rate, an alternative simple choice is also available, namely a coordinate time. In the present context, one could choose the Minkowski coordinate time. Nevertheless, as will become clear throughout the text, a more sensible path is to adopt the Rindler coordinate time $\eta$. Hence, in what follows we describe the time evolution with respect to such a parameter which, because of~(\ref{propt}), has a functional relation with each of the proper times of the atoms.

The stationary trajectories guarantees that the undisturbed atomic system has stationary states which in our particular case are given by equation~(\ref{sta}). Within the multipolar coupling approach the purely atomic part of the total Hamiltonian describes the free Hamiltonian of two two-level atoms. The Hamiltonian that governs the time evolution of this atomic system with respect to $\eta$ can then be written as
\bea
H_A(\eta) &=& \frac{\omega_0}{2}\biggl[\left(S_{1}^z(\tau_1(\eta))\otimes\hat{1}\right)\,\frac{d\tau_1}{d\eta}
\nn\\
&+&\,\left(\hat{1}\otimes S_{2}^z(\tau_2(\eta))\right)\,\frac{d\tau_2}{d\eta}\biggr],
\label{ha}
\eea
where $S_a^z = | e_a \rangle\langle e_a | - | g_a \rangle\langle g_a |$, $a = 1, 2$, $|g_1\rangle$ and $|g_2\rangle$ being the ground states of the atoms isolated from each other and $|e_1\rangle$ and $|e_2\rangle$ their respective excited states. The space of the free two-atom system is spanned by four product states with respective energies
\bea
&& E_{gg} = -\omega_0\,\,\,\,|gg\rangle = |g_1\rangle|g_2\rangle,
\nn\\
&& E_{ge}= 0\,\,\,\,|ge\rangle = |g_1\rangle|e_2\rangle,
\nn\\
&&E_{eg} = 0\,\,\,\,|eg\rangle = |e_1\rangle|g_2\rangle,
\nn\\
&&E_{ee} = \omega_0\,\,\,\,|ee\rangle = |e_1\rangle|e_2\rangle,
\label{sta}
\eea
where a tensor product is implicit. Instead of working with this product-state basis, we can conveniently choose the Bell state basis. In terms of the product states, one has:
\bea
|\Psi^{\pm}\rangle &=& \frac{1}{\sqrt{2}}\left(|g_1\rangle|e_2\rangle \pm |e_1\rangle|g_2\rangle\right)
\nn\\
|\Phi^{\pm}\rangle &=& \frac{1}{\sqrt{2}}\left(|g_1\rangle|g_2\rangle \pm |e_1\rangle|e_2\rangle\right).
\label{bell}
\eea
The Bell states are known as the four maximally entangled two-qubit Bell states, and they form a convenient basis of the two-qubit Hilbert space. Even though we consider identical atoms we are employing the above notation in order to allow the possibility of a straighforward generalization of our results to the case of non-identical atoms. Furthermore, we remark that the Bell states $|\Phi^{\pm}\rangle$ are not eigenstates of the atomic Hamiltonian $H_A$.

Here we consider that the two-atom system is coupled with an electromagnetic field. By exploring the gauge invariance of the electromagnetic Lagrangian, we impose the transversality condition (the Coulomb gauge). Upon the usual decomposition of the electric and magnetic field operators in terms of creation and annihilation operators 
\bea
{\bf E}(t,{\bf x}) &=&\sum_{\lambda} \int d^3 k\,\frac{i\omega_{{\bf k}}}{(2\pi)^3 (2 \omega_{{\bf k}})^{1/2}}\,
\biggl[a_{{\bf k},\lambda}(t)\,e^{i {\bf k}\cdot {\bf x}} 
\nn\\
&-&\, a^{\dagger}_{{\bf k},\lambda}(t)\,e^{-i {\bf k}\cdot {\bf x}}\biggr]{\hat \varepsilon}_{\lambda}({\bf k}),
\eea
\bea
{\bf B}(t,{\bf x}) &=& \sum_{\lambda} \int d^3 k\,\frac{i\omega_{{\bf k}}}{(2\pi)^3 (2 \omega_{{\bf k}})^{1/2}}\,\biggl[a_{{\bf k},\lambda}(t)\,e^{i {\bf k}\cdot {\bf x}} 
\nn\\
&-&\, a^{\dagger}_{{\bf k},\lambda}(t)\,e^{-i {\bf k}\cdot {\bf x}}\biggr]({\bf {\hat k}}\times{\hat \varepsilon}_{\lambda}({\bf k}))
\eea
the electromagnetic field Hamiltonian $H_F = (1/2)\int d^3 x\, \left[{\bf E}^2 + {\bf B}^2\right]$ acquires the traditional form
\beq
H_F(t) = \sum_{\lambda}\int \frac{d^3 k}{(2\pi)^3}\, \omega_{{\bf k}}\,a^{\dagger}_{{\bf k},\lambda}(t)a_{{\bf k},\lambda}(t),
\label{hf}
\eeq
where $a^{\dagger}_{{\bf k},\lambda}, a_{{\bf k},\lambda}$ are the creation and annihilation operators of the electromagnetic field and we have neglected the zero-point energy. The index $\lambda$ labels different polarizations of the field: The orthonormal vectors ${\hat \varepsilon}_{\lambda}({\bf k})$, $\lambda = 1, 2$ correspond to the two polarization states of the photon. These obey known relations that can be found in many textbooks, see for instance~\cite{cohen}. In the multipolar coupling scheme and within the so-called electric-dipole approximation one has that the Hamiltonian which describes the interaction between the atoms and the field is given by
\bea
H_{I}(\eta) &=& - \boldsymbol\mu_1(\tau_1(\eta)) \cdot {\bf E}(x_1(\tau_1(\eta)))\,\frac{d\tau_1}{d\eta}
\nn\\
&-&\, \boldsymbol\mu_2(\tau_2(\eta)) \cdot {\bf E}(x_2(\tau_2(\eta)))\,\frac{d\tau_2}{d\eta}
\label{hi}
\eea
where $\boldsymbol\mu_i$ ($i=1,2$) is the electric dipole moment operator for the $i$-th atom. It should be observed that the electric field above is the measured electric field defined through the measured force it exerts on an atom. The appearence of terms as $d\tau/d\eta$ is a consequence of the fact that both atoms move along different stationary trajectories as discussed above. As asserted in this paper we describe time evolutions with a coordinate time parameter which is to be identified with the Rindler coordiante time $\eta$. Moreover, we are assuming that for both atoms each component of the dipole moment is required to evolve in time in the same way as De Witt's monopole in the case of a scalar field, as discussed by Takagi~\cite{takagi}. This would be the case only if the directions of the both dipole moments are kept fixed with respect to the proper frame of reference of the atoms; otherwise the rotation of the dipole moment will bring in extra time dependence in addition to its intrinsic time evolution. In addition, one should bear in mind that, for each atom $\boldsymbol\mu\cdot {\bf E} = \sum_{j}\mu_{j}\,E^{j}$, where $j$ stands for the $j$-th spatial component with respect to the proper reference frame of the atom, which is the Rindler orthonormal frame of reference. The expression for the atom-field interaction can always be put in a manifestly invariant form, that is, for each atom: $U^{\mu}(\tau)\,F_{\mu\nu}(x(\tau))v^{\nu}(\tau)$, where $v^{\mu}$ is the four-velocity of the atom and $U^{\mu}$ is the four-vector associated with the dipole moment such that, in its proper reference frame, its temporal component vanishes and its spatial components are given by $\mu^{i}$.


We point out that the dipole moment operator can be written as
\bea
\boldsymbol\mu_i(\tau_i) &=& \langle g_i |\boldsymbol\mu_i| e_i \rangle | g_i \rangle\langle e_i| + \textrm{H.c.}
\nn\\
&\equiv&\, \boldsymbol\mu\left[S_{i}^{+}(\tau_i) + S_{i}^{-}(\tau_i)\right],
\label{dip}
\eea
(no summation over repeated indices) where we have assumed a proper choice of phases which allows the dipole matrix elements to be real and we have defined the dipole raising and lowering operators as $S_{i}^{+} = | e_i \rangle\langle g_i|$ and $S_{i}^{-} = | g_i \rangle\langle e_i|$, respectively. In addition, since the atoms are presumed to be identical, the matrix elements $\langle g_i |\boldsymbol\mu_i| e_i \rangle$ are independent of the index $i$ and denoted simply by $\boldsymbol\mu$.

The Heisenberg equations of motion for the dynamical variables of the atom and the field can be derived from the total Hamiltonian 
$H = H_A + H_F + H_I$. We draw attention to the fact that the present work employs a coherent quantum evolution treatment instead of the more tradicional open quantum system approach in the analysis of quantum entanglement. As remarked in the previous Section, the former allows one to discuss the contributions of vacuum fluctuations and radiation reaction in the radiative processes of entangled atoms. The rates of change of the atomic observables are governed by the Heisenberg equations
\bea
\frac{d S_a^z(\tau_a(\eta))}{d\eta}&=&  - i \,[\boldsymbol\mu_1(\tau_1(\eta))\cdot{\bf E}(x_1(\tau_1(\eta))),S_a^z(\tau_a(\eta))]\,\frac{d\tau_1}{d\eta}
\nn\\
&-&\, i\,[\boldsymbol\mu_2(\tau_2(\eta))\cdot{\bf E}(x_2(\tau_2(\eta))),S_a^z(\tau_a(\eta))]\,\frac{d\tau_2}{d\eta},
\nn\\
\label{bad1}
\eea
$a = 1, 2$, whereas for dynamical variables of the field one has
\bea
\frac{d}{dt}a_{{\bf k},\lambda}(t(\eta))\, &&= -i\omega_{{\bf k}}a_{{\bf k},\lambda}(t(\eta))
\nn\\
&&- i[\boldsymbol\mu_1(\tau_1)\cdot{\bf E}(x_1(\tau_1)),a_{{\bf k},\lambda}(t(\eta))]\frac{d\tau_1}{dt}
\nn\\
&&- i[\boldsymbol\mu_2(\tau_2)\cdot{\bf E}(x_2(\tau_2)),a_{{\bf k},\lambda}(t(\eta))]\frac{d\tau_2}{dt}.\nn\\
\label{bad2}
\eea
In Eq.~(\ref{bad2}), $t$ and the proper times $\tau_a$ must be considered as functions of $\eta$. In addition, in Eq.~(\ref{bad1}) one have taken into account that $S_a^z$ commutes with $H_A$. In order to proceed one splits the solutions of the equations of motion in two parts, namely: The free part, which is present even in the absence of the coupling; and the source part, which is caused by the interaction between atom and field and contains the coupling constant $e$. Therefore one can write~\cite{aud1,passante}
\bea
S_a^z(\tau_a(\eta)) &=& S_a^{z,f}(\tau_a(\eta)) + S_a^{z,s}(\tau_a(\eta)),
\nn\\
a_{{\bf k},\lambda}(t(\eta)) &=& a^{f}_{{\bf k},\lambda}(t(\eta)) + a^{s}_{{\bf k},\lambda}(t(\eta)).
\eea
Similar considerations can be applied for the the electric dipole moment operators of the atoms, namely: $\boldsymbol\mu_a^z(\tau_a(\eta)) = \boldsymbol\mu_a^{z,f}(\tau_a(\eta)) + \boldsymbol\mu_a^{z,s}(\tau_a(\eta))$. In this way, up to first order in $\mu$, the solutions of the equations of motion are given by:
\begin{widetext}
\bea
S_a^{z,f}(\tau_a(\eta)) &=& S_a^{z,f}(\tau_a(\eta_0))
\nn\\
S_a^{z,s}(\tau_a(\eta)) &=& - i \,\sum_{b =1}^{2}\int_{\eta_0}^{\eta}\,d\eta' [\boldsymbol\mu^{f}_b(\tau_b(\eta'))
\cdot{\bf E}^{f}(x_b(\tau_b(\eta'))),S_a^{z,f}(\tau_a(\eta))]\,\frac{d\tau_b}{d\eta'}
\nn\\
a^{f}_{{\bf k},\lambda}(t(\eta)) &=&  a^{f}_{{\bf k},\lambda}(t(\eta_0))\,e^{-i\omega_{{\bf k}} [t(\eta) - t(\eta_0)]}
\nn\\
a^{s}_{{\bf k},\lambda}(t(\eta)) &=& -i\sum_{b =1}^{2}\int_{\eta_0}^{\eta}\,d\eta'[\boldsymbol\mu^{f}_b(\tau_b(\eta'))
\cdot{\bf E}^{f}(x_b(\tau_b(\eta'))),a^{f}_{{\bf k},\lambda}(t(\eta))]\frac{d\tau_b}{d\eta'}.
\eea
\end{widetext}
Since one can construct from the annihilation and creation field operators the free and source part of the quantum electric field, one also has
\beq
{\bf E}(t(\eta),{\bf x}(\eta)) = {\bf E}^{f}(t(\eta),{\bf x}(\eta)) + {\bf E}^{s}(t(\eta),{\bf x}(\eta)).
\eeq
However, as discussed in Refs.~\cite{cohen2,cohen3,aud1}, there is an ambiguity of operator ordering which arises in such a procedure. Being more specific, when one considers the Heisenberg equations of motion for an arbitrary atomic observable $A(\tau)$
\beq
\frac{d A}{d\eta} = \frac{1}{i}\,[A(\eta), H_{A}(\eta)] + \frac{1}{i}\,[A(\eta), H_{I}(\eta)]
\eeq
and proceeds with the split of the solutions for the quantum field into free and source parts. The feature that all atomic observables commute with ${\bf E}$ is preserved in time because of the unitary evolution. However, this is no longer true for ${\bf E}^{f}$ and ${\bf E}^s$ separately: the source part of the field picks up contributions of atomic observables during its time evolution. In face of this fact, one must choose an operator ordering when discussing the effects of ${\bf E}^{f}$ and ${\bf E}^s$ separately. This operation does not modify the final results for physical quantities, yet it does produce distinct interpretations concerning the roles played by vacuum fluctuations and radiation reaction. Nonetheless, there exists a preferred operator ordering by which the effects of such phenomena can posses independent physical meanings~\cite{cohen2,cohen3,aud1}. In other words, such works reveal that only if a symmetric ordering of atomic and field variables is adopted, both contributions coming from vacuum fluctuations and radiation reaction are Hermitian. Therefore, adopting the symmetric ordering prescription, one can identify the contribution of the vacuum fluctuations to the above rate:
\begin{widetext}
\bea
\left(\frac{d A}{d\eta}\right)_{VF} &=& - \frac{i}{2} \,\sum_{a=1}^{2}
\Bigl\{{\bf E}^{f}(x_a(\tau_a(\eta)))\cdot[\boldsymbol\mu_a(\tau_a(\eta)),A(\eta)]
+ [\boldsymbol\mu_a(\tau_a(\eta)),A(\eta)]\cdot{\bf E}^{f}(x_a(\tau_a(\eta)))\Bigr\}\,\frac{d\tau_a}{d\eta},
\label{vf}
\eea
where the subscript $VF$ stands for the vacuum-fluctuation contribution, and the radiation reaction terms read
\bea
\left(\frac{d A}{d\eta}\right)_{RR} &=& - \frac{i}{2} \,\sum_{a=1}^{2}
\Bigl\{{\bf E}^{s}(x_a(\tau_a(\eta)))\cdot[\boldsymbol\mu_a(\tau_a(\eta)),A(\eta)]
+ [\boldsymbol\mu_a(\tau_a(\eta)),A(\eta)]\cdot{\bf E}^{s}(x_a(\tau_a(\eta)))\Bigr\}\,\frac{d\tau_a}{d\eta},
\label{rr}
\eea
with the subscript $RR$ labeling the radiation-reaction effect terms. In the expressions above we have considered only the part due to the interactions with the field. This method is the most interesting instruction which allows the possibility of studying the interplay between vacuum fluctuations and radiation reaction in quantum entanglement. Hence one can identify the contributions of vacuum fluctuations and radiation reaction in the evolution of the atoms'  energies, which are given by the expectation value of $H_A$, given by equation~(\ref{ha}). The free part of the atomic Hamiltonian is constant in time so that the rate of change of $H_A$ consists only of the two contributions given by, respectively
\bea
\left(\frac{d H_A}{d\eta}\right)_{VF} &=& - \frac{i}{2} \,\sum_{a=1}^{2}
\Bigl\{{\bf E}^{f}(x_a(\tau_a(\eta)))\cdot[\boldsymbol\mu_a(\tau_a(\eta)),H_A(\eta)]
+ [\boldsymbol\mu_a(\tau_a(\eta)),H_A(\eta)]\cdot{\bf E}^{f}(x_a(\tau_a(\eta)))\Bigr\}\,\frac{d\tau_a}{d\eta},
\label{vfha}
\eea
and
\bea
\left(\frac{d H_A}{d\eta}\right)_{RR} &=& - \frac{i}{2} \,\sum_{a=1}^{2}
\Bigl\{{\bf E}^{s}(x_a(\tau_a(\eta)))\cdot[\boldsymbol\mu_a(\tau_a(\eta)),H_A(\eta)]
+ [\boldsymbol\mu_a(\tau_a(\eta)),H_A(\eta)]\cdot{\bf E}^{s}(x_a(\tau_a(\eta)))\Bigr\}\,\frac{d\tau_a}{d\eta}.
\label{rrha}
\eea
In a perturbative treatment, we take into account only terms up to order $\mu^2$. We also consider an averaging over the field degrees of freedom by taking vacuum expectation values. Since they contain only free operators, only the electric field is affected. In turn, we are interested in the evolution of expectation values of atomic observables. Accordingly, we also take the expectation value of the above expressions in a state $|\omega\rangle$ with energy $\omega$, which can be one of the states given by equation~(\ref{sta}) or one can also consider the Bell states $|\Psi^{\pm}\rangle$. More details concerning such calculations can be found elsewhere, see for instance~\cite{aud1,passante}. Therefore, with the notation $\langle (\cdots) \rangle = \langle 0,\omega |(\cdots)| 0,\omega \rangle$, one has the pivotal results concerning the vacuum-fluctuation contribution
\beq
\Biggl\langle \frac{d H_A}{d\eta} \Biggr\rangle_{VF} = \frac{i}{2}\int_{\eta_0}^{\eta}d\eta' \,\sum_{a,b = 1}^{2}\,\frac{d\tau_a}{d\eta}\frac{d\tau_b'}{d\eta'}\,D_{ij}(x_a(\tau_a(\eta)),x_b(\tau_b'(\eta')))\frac{\partial}{\partial\tau_a}\Delta^{ij}_{ab}(\tau_a(\eta),\tau_b'(\eta')),
\label{vfha3}
\eeq
where
\beq
\Delta^{ij}_{ab}(\tau_a(\eta),\tau_b'(\eta')) = \langle \omega | [\mu^{i,f}_{a}(\tau_a(\eta)),\mu^{j,f}_{b}(\tau_b'(\eta'))]| \omega \rangle,\,\,\,a,b = 1,2,
\label{susa}
\eeq
is the linear susceptibility of the two-atom system in the state $|\omega\rangle$ and
$$D_{ij}(x_a(\tau_a(\eta)),x_b(\tau_b'(\eta'))) = \langle 0 |\{E^{f}_{i}(x_a(\tau_a(\eta))),E^{f}_{j}(x_b(\tau_b'(\eta')))\}| 0 \rangle,\,\,\,a, b =1,2
$$
is the Hadamard's elementary function. Since we are dealing with free fields, one can employ the results derived in the Appendix~\ref{A}. On the other hand, for the radiation-reaction contribution, one has:
\beq
\Biggl\langle \frac{d H_A}{d\eta} \Biggr\rangle_{RR} = \frac{i}{2}\int_{\eta_0}^{\eta}d\eta' \,\sum_{a,b=1}^{2}\,\frac{d\tau_a}{d\eta}\frac{d\tau_b'}{d\eta'}\,\Delta_{ij}(x_a(\tau_a(\eta)),x_b(\tau_b'(\eta')))\frac{\partial}{\partial\tau_a} D^{ij}_{ab}(\tau_a(\eta),\tau_b'(\eta')),
\label{rrha3}
\eeq
where
\beq
D^{ij}_{ab}(\tau_a(\eta),\tau_b'(\eta')) = \langle \omega | \{\mu^{i,f}_{a}(\tau_a(\eta)),\mu^{j,f}_{b}(\tau_b'(\eta'))\}| \omega \rangle,\,\,\,a,b = 1,2,
\label{cora}
\eeq
is the symmetric correlation function of the two-atom system in the state $|\omega\rangle$ and
$$\Delta_{ij}(x_a(\tau_a(\eta)),x_b(\tau_b'(\eta'))) = \langle 0 |[E^{f}_{i}(x_a(\tau_a(\eta))),E^{f}_{j}(x_b(\tau_b'(\eta')))]| 0 \rangle,\,\,\,a, b =1,2
$$
is the Pauli-Jordan function which, for the same reason as above, can be evaluated from Appendix~\ref{A}. Observe that we have employed an index notation in order to represent the dot product of two vectors ${\bf E}\cdot\boldsymbol\mu = E_{i}\,\mu^{i}$, where the Einstein summation convention is to be understood in the above equations and henceforth (unless otherwise stated). Therefore the vector character of the electric field is manifest in the appearance of indices in the correlation functions.
\end{widetext}

Note that $\Delta^{ij}_{ab}$ and $D^{ij}_{ab}$ characterize only the two-atom system itself, unlike the statistical functions of the field which have to be evaluated along the trajectory of the atoms. We see from equations~(\ref{vfha3}) and~(\ref{rrha3}) that the rate of variation of the energy of the two-atom system presents contributions from the isolated atoms and also contributions due to cross correlations between the atoms mediated by the field. This interference is a consequence of the interaction
of each atom with the field. Such observations make clear that such a formalism enables one to discuss the generation of entanglement between atoms initially prepared, say, in the ground state as well as the stability of entangled states.

Using that
$$
\mu^{i,f}_{a}(\tau_a) = e^{i H_A\tau_a}\mu^{i,f}_{a}(0)e^{-i H_A\tau_a},
$$
$a = 1, 2$, and the completeness relation for the atomic product states (or the Bell states):
$$
\sum_{\omega'}| \omega' \rangle \langle \omega' | = {\bf 1},
$$
one can show that the statistical functions for the two-atom system can be put in the following explicit forms
\bea
\Delta^{ij}_{ab}(\eta,\eta') &=& \sum_{\omega'} \biggl[{\cal A}^{ij}_{ab}(\omega,\omega')\,e^{i\Delta\omega(\tau_a(\eta) - \tau_b(\eta'))}
\nn\\
&-&\,{\cal A}^{ji}_{ba}(\omega,\omega')\,e^{-i\Delta\omega(\tau_a(\eta) - \tau_b(\eta'))}\biggr],
\label{susa1}
\eea
and
\bea
D^{ij}_{ab}(\eta,\eta') &=& \sum_{\omega'} \biggl[{\cal A}^{ij}_{ab}(\omega,\omega')
\,e^{i\Delta\omega(\tau_a(\eta) - \tau_b(\eta'))}
\nn\\
&+&\, {\cal A}^{ji}_{ba}(\omega,\omega')\,e^{-i\Delta\omega(\tau_a(\eta) - \tau_b(\eta'))}\biggr],
\label{cora1}
\eea
where $\Delta\omega = \omega - \omega'$, $\omega'$ being the energy associated with the state $|\omega'\rangle$. If the initial state $|\omega\rangle$ is one of the Bell entangled states, then the sum over $\omega'$ is carried out with respect to the product states given by Eq.~(\ref{sta}). On the other hand, if the initial state $|\omega\rangle$ is one of the states of Eq.~(\ref{sta}), then such a sum is over the Bell states of Eq.~(\ref{bell}). In the above the generalized atomic transition dipole moment ${\cal A}^{ij}_{cd}(\omega,\omega')$ is defined by
\beq
{\cal A}^{ij}_{cd}(\omega,\omega') =  \langle \omega |\mu^{i,f}_{c}(0)| \omega' \rangle\langle \omega' |\mu^{j,f}_{d}(0)| \omega \rangle.
\label{aa}
\eeq
Finally, observe that from~(\ref{propt}) one can easily perform a change of variables in equations~(\ref{vfha3}) and~(\ref{rrha3}) in order to describe the time evolution in terms of one of the proper times of the atoms.

In this paper we wish to investigate the creation of entangled states [currently represented by the Bell states~(\ref{bell})] from atoms initially prepared in a separable state [given by equation~(\ref{sta})], and also the decay of the Bell states to one of the product states. In order to study the degradation of entanglement between the atoms as a spontaneous emission phenomenon, assume that the atoms were initially prepared in one of the Bell states $|\Psi^{\pm}\rangle$. With this regard, equations~(\ref{sta}) and~(\ref{bell}) state that the only allowed transitions are $|\Psi^{\pm}\rangle \to |gg\rangle$, with $\Delta \omega = \omega - \omega' = \omega_0 > 0$ and $|\Psi^{\pm}\rangle \to |ee\rangle$, with $\Delta \omega = \omega - \omega' = - \omega_0 < 0$. On the other hand, suppose one wishes to address the generation of entanglement. Assume that the atoms were initially prepared in the atomic ground state $|gg\rangle$.  Consider the excitation rate to one of the Bell states $|\Psi^{\pm}\rangle$. Hence $\Delta \omega = - \omega_0 < 0$. Similarly, for atoms initially prepared in the state $|ee\rangle$, only the transition $|ee\rangle \to |\Psi^{\pm}\rangle$ is permitted, with $\Delta \omega = \omega - \omega' = \omega_0 > 0$. Incidentally, we remark that, if we consider the excitation rate to one of the Bell states $|\Phi^{\pm}\rangle$, one would get $\Delta \omega = 0$, which implies that $\langle d H_A/d\eta \rangle = 0$. Hence it is not possible to generate such Bell states out of the ground-state atoms.

In the next Section we will discuss in detail the rate of variation of atomic energy for atoms in uniformly accelerated motion.

\section{Rate of variation of the atomic energy in vacuum for uniformly accelerated motion}
\label{atom}

\subsection{Presentation of the main results}

We consider our two-atom system in a situation where both atoms move along different hyperbolic trajectories in the $(t,x_1)$ plane. As explained, these correspond to different $\xi$ coordinates, which we call $\xi_1$ and $\xi_2$ and each $\xi$ is a constant. Their world lines are parametrized by equations~(\ref{acc}). For simplicity, we consider the case in which both atoms have the same spatial coordinates which are orthogonal to the direction of the acceleration. We meticulously evaluate the expressions concerning the contributions of vacuum fluctuations and radiation reaction to the rate of variation of atomic energy in Appendix~\ref{B}. Here we only present the main results. Concerning the contributions from vacuum fluctuations, one has:
\begin{widetext}
\bea
&& \Biggl\langle \frac{d H_A}{d\eta} \Biggr\rangle_{VF} =  -\frac{a^3}{4\pi}\sum_{\omega'}\sum_{c,d = 1}^{2}e^{- a(\xi_c + \xi_d)}\,\exp\left[\frac{i\Delta\omega(\alpha_d -\alpha_c)a\eta}{\alpha_c\alpha_d}\right]\,\text{csch}^2\left(\psi_{cd}\right)
\nn\\
&\times&\,\Biggl\{\Delta\omega\,{\cal J}^{\omega\omega'}_{cd}(\Delta\omega,\Delta\eta)
+\theta(\Delta\omega)\Delta\omega\,\delta_{ij}\Biggl[{\cal F}^{ij}_{1,cd}(\Delta\omega,\alpha_d)
+\frac{{\cal F}^{ij}_{2,cd}(\Delta\omega,\alpha_d)}{e^{2\pi \Delta\omega /\alpha_d} - 1} + \Bigl({\cal F}^{ij}_{1,cd}(\Delta\omega,\alpha_c)\Bigr)^{*}
+\frac{\Bigl({\cal F}^{ij}_{2,cd}(\Delta\omega,\alpha_c)\Bigr)^{*} }
{e^{2\pi \Delta\omega /\alpha_c} - 1}\Biggr]
\nn\\
&-&\, \theta(-\Delta\omega)|\Delta\omega|\,\delta_{ij}\Biggl[\Bigl({\cal F}^{ij}_{1,cd}(|\Delta\omega|,\alpha_d)\Bigr)^{*}
+\frac{\Bigl({\cal F}^{ij}_{2,cd}(|\Delta\omega|,\alpha_d)\Bigr)^{*} }
{e^{2\pi |\Delta\omega| /\alpha_d} - 1} + {\cal F}^{ij}_{1,cd}(|\Delta\omega|,\alpha_c)
+\frac{{\cal F}^{ij}_{2,cd}(|\Delta\omega|,\alpha_c)}{e^{2\pi |\Delta\omega| /\alpha_c} - 1}\Biggr]\Biggr\},
\label{vacuum2}
\eea
where $\Delta\eta > \psi_{cd}/a > 0$. We have defined the following functions:
\bea
{\cal J}^{\omega\omega'}_{cd}(\Delta\omega, \Delta\eta) &=& - \frac{1}{8\pi\text{csch}^2\psi_{cd}}
\int_{a\Delta\eta}^{\infty}du \,\left(e^{i\Delta\omega u/\alpha_d} + e^{-i\Delta\omega u/\alpha_c}\right)
\,\frac{\delta_{ij} T^{ij}_{cd}(\omega,\omega'; u, 0)}{\sinh^3\left(\frac{a u + \psi_{cd}}{2} \right)\sinh^3\left(\frac{a u - \psi_{cd}}{2} \right)}
\label{rev28}
\eea
\end{widetext}
and
\bea
{\cal F}^{ij}_{n,cd}(\Delta\omega,\alpha) &=& {\cal F}^{ij}_{n,cd}(\omega, \omega'; \Delta\omega,\alpha)
\nn\\
&=& n\,U^{ij}_{cd}(\omega,\omega'; \Delta\omega,\alpha) + V^{ij}_{cd}(\omega, \omega'; \Delta\omega,\alpha)
\nn\\
\label{rev29}
\eea
where we also have introduced the following definitions:
\bea
U^{ij}_{cd}(\Delta\omega,\alpha) &=& U^{ij}_{cd}(\omega,\omega'; \Delta\omega,\alpha)
\nn\\
&=& {\cal A}^{ij}_{cd}(\omega,\omega') {\cal U}_{i,cd}(\Delta\omega,\alpha)\,
\label{rev30}
\eea
with no implicit summation on repeated indices, and
\bea
V^{ij}_{cd}(\Delta\omega, \alpha) &=& V^{ij}_{cd}(\omega, \omega'; \Delta\omega, \alpha)
\nn\\
&=&\frac{i}{\pi\text{csch}^2\psi_{cd}}\int_{0}^{2\pi}du \,e^{-\Delta\omega u/\alpha}
\nn\\
&\times&\,\frac{T^{ij}_{cd}(\omega,\omega'; iu, 0)}{\left(\cos u - \cosh\psi_{cd} \right)^3}.
\label{rev31}
\eea
The function $\psi_{cd}$ is given in the Appendix~\ref{A} whilst the functions $T^{ij}_{cd}(\omega,\omega'; u,\epsilon)$ and ${\cal U}_{i,cd}(\Delta\omega,\alpha)$ are properly defined in the Appendix~\ref{B}, see equations~(\ref{b2}),~(\ref{d1}) and~(\ref{d23}), respectively. Now let us present the radiation-reaction contributions. One gets, with $\Delta\eta > \psi_{cd}/a > 0$:
\begin{widetext}
\bea
\Biggl\langle \frac{d H_A}{d\eta} \Biggr\rangle_{RR} &=& -\frac{a^3}{4\pi}\sum_{\omega'}\sum_{c,d = 1}^{2}e^{- a(\xi_c + \xi_d)}\,\exp\left[\frac{i\Delta\omega(\alpha_d -\alpha_c)a\eta}{\alpha_c\alpha_d}\right]\,\text{csch}^2\left(\psi_{cd}\right)
\nn\\
&\times&\,\Biggl\{\theta(\Delta\omega)\Delta\omega\Biggl[\delta_{ij}U^{ij}_{cd}(\Delta\omega,\alpha_d)
+\delta_{ij}\Bigl(U^{ij}_{cd}(\Delta\omega,\alpha_c)\Bigr)^{*}\Biggr]
\nn\\
&+&\, \theta(-\Delta\omega)|\Delta\omega|\Biggl[\delta_{ij}\Bigl(U^{ij}_{cd}(|\Delta\omega|,\alpha_d)\Bigr)^{*}
+\delta_{ij}U^{ij}_{cd}(|\Delta\omega|,\alpha_c) \Biggr]\Biggr\}.
\label{radiation2}
\eea
\end{widetext}
The prior consideration to be taken regarding the above results is that the effects of electromagnetic vacuum fluctuations over the rate of variation of energy of an accelerated atom are not purely thermal: there exist nonthermal corrections proportional to the acceleration of the atoms. A similar result is found in reference~\cite{china1}, where the authors study a multilevel hydrogen atom in interaction with the quantum electromagnetic field. With respect to the radiation-reaction effect, we also find a striking dependence on the acceleration. Again this situation agrees with the result found in reference~\cite{china1}.

For completeness we present the total rate of change of the energy of the two-atom system in a general situation. This is obtained by adding the contributions of vacuum fluctuations and radiation reaction given by equations~(\ref{vacuum2}) and~(\ref{radiation2}), respectively. We assume sufficiently large observation time intervals $\Delta\eta$ in such a way that the term ${\cal J}^{\omega\omega'}_{cd}(\Delta\omega, \Delta\eta)$ can be neglected. One has:
\begin{widetext}
\bea
 \Biggl\langle \frac{d H_A}{d\eta} \Biggr\rangle_{tot} &=&  -\frac{a^3}{4\pi}\sum_{\omega'}\sum_{c,d = 1}^{2}e^{- a(\xi_c + \xi_d)}\,\exp\left[\frac{i\Delta\omega(\alpha_d -\alpha_c)a\eta}{\alpha_c\alpha_d}\right]\,\text{csch}^2\left(\psi_{cd}\right)\,\Delta\omega
\nn\\
&\times&\delta_{ij}\,\Biggl[{\cal F}^{ij}_{2,cd}(\Delta\omega,\alpha_d)\left(1
+\frac{1}{e^{2\pi \Delta\omega /\alpha_d} - 1}\right) + \Bigl({\cal F}^{ij}_{2,cd}(\Delta\omega,\alpha_c)\Bigr)^{*}\left(1  +
\frac{1}{e^{2\pi \Delta\omega /\alpha_c} - 1}\right)\Biggr],
\eea
with $\Delta\omega > 0 $, and
\bea
\Biggl\langle \frac{d H_A}{d\eta} \Biggr\rangle_{tot} &=& \frac{a^3}{4\pi}\sum_{\omega'}\sum_{c,d = 1}^{2}e^{- a(\xi_c + \xi_d)}\,\exp\left[\frac{-i|\Delta\omega|(\alpha_d -\alpha_c)a\eta}{\alpha_c\alpha_d}\right]\,\text{csch}^2\left(\psi_{cd}\right)\,
|\Delta\omega|
\nn\\
&\times&\delta_{ij}\,\Biggl[\Bigl(V^{ij}_{cd}(|\Delta\omega|,\alpha_d)\Bigr)^{*} + V^{ij}_{cd}(|\Delta\omega|,\alpha_c)
+\frac{\Bigl({\cal F}^{ij}_{2,cd}(|\Delta\omega|,\alpha_d)\Bigr)^{*}}{e^{2\pi |\Delta\omega| /\alpha_d} - 1}
+\frac{{\cal F}^{ij}_{2,cd}(|\Delta\omega|,\alpha_c)}{e^{2\pi |\Delta\omega| /\alpha_c} - 1}\Biggr],
\eea
\end{widetext}
with $\Delta\omega < 0$. Notice that in the ground state of the two-atom system there is not a balance between vacuum fluctuations and radiation reaction which prevents spontaneous excitations to higher levels. In our context this means that entanglement between the atoms can be generated via absorption processes and it persists even for asymptotic observation times. Observe also that the cross correlations introduce strong temporal oscillations.

In order to carefully discuss the thermal contributions, let us consider the following situation. Imagine a family of particle detectors at rest in different points in Rindler space-time. Even though each detector measures different temperatures $T(x)$, one says that all of them are in thermal equilibrium due to the Tolman relation $\sqrt{g_{00}(x)}\,T(x) = \textrm{const.}$~\cite{tolman}. We find a similar structure here, namely rates possessing thermal contributions containing each of the temperatures felt by the atoms. By invoking the thermalization theorem~\cite{sciama,fulling,davies}, one concludes that in the situation with sufficiently high relative accelerations we have two atoms coupled individually to two spatially separated cavities at different temperatures. Hence our results clearly indicates that such a situation bears a remarkable resemblance with the initial set up of Ref.~\cite{eberly}. We infer that if a detailed investigation of a circumstance with high relative accelerations is carried out within a master equation approach, one should expect to obtain similar results as uncovered by the authors of such a reference. Namely, that complete disentanglement can be achieved after finite time processes, yet local decoherence operations requires an infinite time to be accomplished. Despite its importance, this interesting analysis is out of the scope of the present work. On the other hand, for a careful recent discussion on entanglement in a quantum system at finite temperature we refer the reader the reference~\cite{hu4}.

Now let us discuss in detail the distinct contributions to the rate of variation of atomic energy for a specific transition.

\subsection{Discussions}

Here we analyze the distinct contributions to the rate of variation of atomic energy associated with the transition
$|gg\rangle \to |\Psi^{\pm}\rangle$. This means that we are investigating the creation of entanglement via absorption processes.  Hence in this context $\Delta\omega = -\omega_0 <0$ in equations~(\ref{vacuum2}) and~(\ref{radiation2}). For the sake of the argument, we consider the generation of the symmetric Bell state $|\Psi^{+}\rangle$ but it is possible to carry out a similar analysis for the antisymmetric Bell state $|\Psi^{-}\rangle$. We assume sufficiently large observation time intervals $\Delta\eta$ in such a way that the term ${\cal J}^{\omega\omega'}_{cd}(\Delta\omega, \Delta\eta)$ can be neglected. It is convenient to write out the expressions in a more explicit way. Concerning the vacuum-fluctuation contribution, one has
\begin{widetext}
\bea
&&\Biggl\langle \frac{d H_A}{d\eta} \Biggr\rangle_{VF} = \frac{\omega_0^4\,\boldsymbol\mu^2}{12\pi}\sum_{c=1}^{2}\,e^{a\xi_c}\,
\left(\frac{\alpha_c^2}{\omega_0^2}+1\right)\left(1+\frac{2}{e^{2\pi \omega_0/\alpha_c} - 1}\right)
\nn\\
&&+\frac{\omega_0\,a^3\,e^{- a(\xi_1 + \xi_2)}}{4\pi}\,\text{csch}^2\left(\psi_{12}\right)\,\delta_{ij}
\Biggl\{{\cal \widetilde{U}}^{i,12}(\omega_0,\alpha_1, \eta) \mu^{j}\left(1+\frac{2}{e^{2\pi \omega_0/\alpha_1} - 1}\right)
+{\cal \widetilde{U}}^{i,21}(\omega_0,\alpha_2, \eta) \mu^{j}\left(1+\frac{2}{e^{2\pi \omega_0/\alpha_2} - 1}\right)
\nn\\
&&+\left[{\cal \widetilde{V}}^{i}_{12}(\omega_0,\alpha_2)\left(1+\frac{1}{e^{2\pi \omega_0/\alpha_2} - 1}\right) - {\cal \widetilde{V}}^{i}_{12}(\omega_0,\alpha_1)\left(1+\frac{1}{e^{2\pi \omega_0/\alpha_1} - 1}\right)\right]\,\mu^{j}\,\sin\left[\frac{\omega_0(\alpha_2 -\alpha_1)a\eta}{\alpha_1\alpha_2}\right]\Biggr\},
\eea
whereas the radiation-reaction terms are given by
\beq
\Biggl\langle \frac{d H_A}{d\eta} \Biggr\rangle_{RR} = -\frac{\omega_0^4\,\boldsymbol\mu^2}{12\pi}\sum_{c=1}^{2}\left(\frac{\alpha_c^2}{\omega_0^2}+1\right)\,e^{a\xi_c}
-\frac{\omega_0\,a^3\,e^{- a(\xi_1 + \xi_2)}}{4\pi}\text{csch}^2\left(\psi_{12}\right)\delta_{ij}
\Biggl[{\cal \widetilde{U}}^{i,12}(\omega_0,\alpha_1, \eta) \mu^{j} + {\cal \widetilde{U}}^{i,21}(\omega_0,\alpha_2, \eta) \mu^{j}\Biggr].
\eeq
In the above expressions we have defined:
\beq
{\cal \widetilde{U}}_{1,cd}(\omega,\alpha_c, \eta) = \mu_1\left\{\sin \left[\frac{\omega \psi_{cd}}{\alpha_c}+\frac{\omega(\alpha_d -\alpha_c)a\eta}{\alpha_c\alpha_d}\right]\coth\psi_{cd}
- \frac{\omega}{\alpha_c}  \cos \left[\frac{\omega \psi_{cd}}{\alpha_c}+\frac{\omega(\alpha_d -\alpha_c)a\eta}{\alpha_c\alpha_d}\right]\right\}
\label{dd1}
\eeq
\bea
 {\cal \widetilde{U}}_{k,cd}(\omega,\alpha_c, \eta) &=&  \frac{\mu_k}{2}\,\cos \left[\frac{\omega \psi_{cd}}{\alpha_c}+\frac{\omega(\alpha_d -\alpha_c)a\eta}{\alpha_c\alpha_d}\right]
\nn\\
&\times&\,\left\{\frac{\omega}{\alpha_c} \cosh \psi_{cd}
+\text{csch}\psi_{cd}\left( \left(\frac{\omega}{\alpha_c}\right)^2 \sinh^2(\psi_{cd})  - 1\right) \tan \left[\frac{\omega \psi_{cd}}{\alpha_c}+\frac{\omega(\alpha_d -\alpha_c)a\eta}{\alpha_c\alpha_d}\right]\right\},
\nn\\
\label{dd23}
\eea
$k = 2, 3$, and
\beq
{\cal \widetilde{V}}_{i,12}(\omega, \alpha) = \frac{\mu_i}{\pi\text{csch}^2\psi_{12}}\int_{0}^{2\pi}du \,e^{-\omega u/\alpha}
\,\frac{\nu_{i,12}(iu, 0)}{\left(\cos u - \cosh\psi_{12} \right)^3},
\label{ddef-v}
\eeq
\end{widetext}
with no summation over the repeated index $i$ in this last expression. The functions $\nu_{i,12}$ are defined in the Appendix~\ref{A}, with the property that $\nu^{*}_{i,12}(u, 0) = \nu_{i,12}(u, 0)$.  Also, we have used the fact that the matrix elements are given by
\bea
{\cal A}^{ij}_{11}(g, \Psi^{\pm}) &=& \frac{\mu^{i}\mu^{j}}{2}
\nn\\
{\cal A}^{ij}_{22}(g, \Psi^{\pm}) &=& \frac{\mu^{i}\mu^{j}}{2}
\nn\\
{\cal A}^{ij}_{12}(g, \Psi^{\pm}) &=& [{\cal A}_{21}(g, \Psi^{\pm})]^{*} = \pm\, \frac{\mu^{i}\mu^{j}}{2}.
\label{mel}
\eea
\begin{figure}[htb]
\begin{center}
\includegraphics[height=80mm,width=88mm]{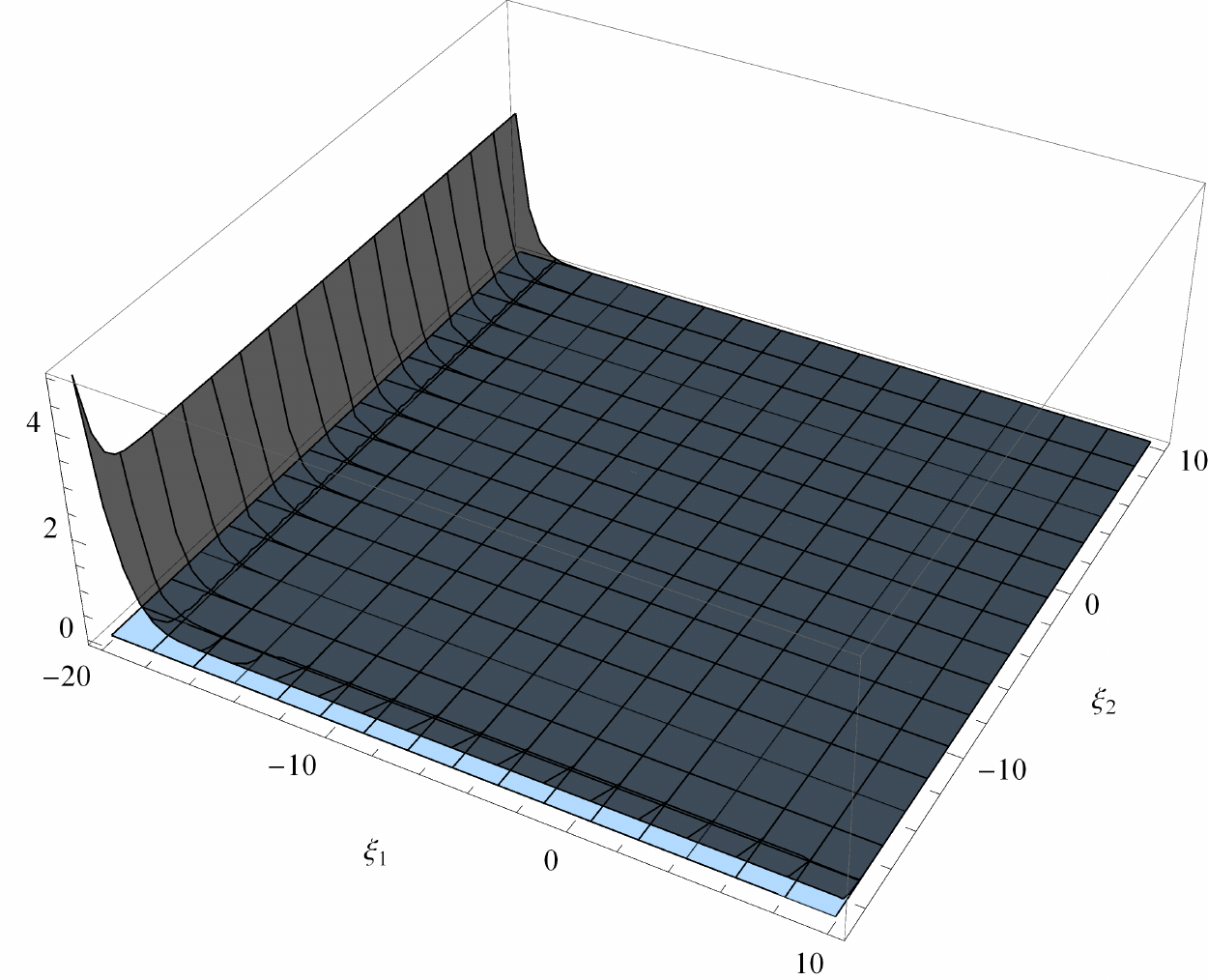}
\caption{Vacuum-fluctuation (in black color) and radiation-reaction (in blue-white color) contributions to the rate of variation of atomic energy as functions of the Rindler coordinates $\xi_1$ and $\xi_2$. We choose  $a=1$, $\omega_0 = 5$, $\eta = 100$ and $\mu^{1} = \mu^{2} = \mu^{3} = 1$. We remark that all physical quantities are given in terms of the natural units associated with each transition. Therefore, in this case $\xi_1$, $\xi_2$, $\eta$ and $\mu$ are measured in units of $\lambda$, $a$ are in units of $\lambda^{-1}$ and $\omega_0$ is given in units of $2\pi\lambda^{-1}$, where $\lambda = 2\pi/\omega_0$. Moreover, $\langle d H_A/d\eta \rangle$ is measured in units of $\lambda^{-2}$. For clarity we have multiplied each value of the rate by $10^{-16}$.}
\label{1}
\end{center}
\end{figure}
\begin{figure}[htb]
\begin{center}
\includegraphics[height=80mm,width=80mm]{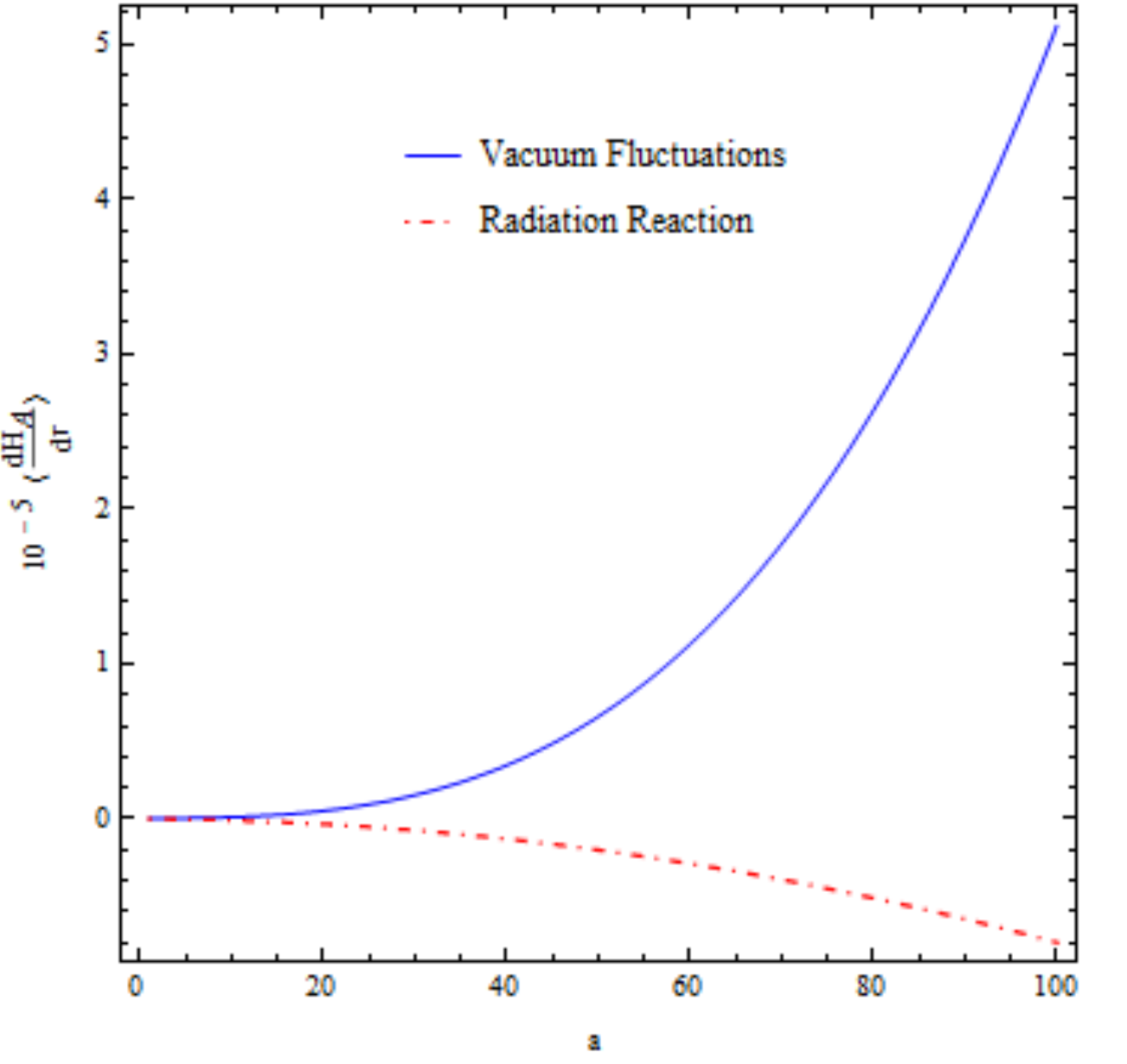}
\caption{Vacuum-fluctuation and radiation-reaction contributions to the rate of variation of atomic energy as functions of the parameter $a$. We consider the atoms with the same proper acceleration. We choose $\omega_0 = 5$, $\mu^{1} = \mu^{2} = \mu^{3} = 1$ and $\xi_1 = \xi_2 = 0$ in natural units associated with the transition in study (see caption below figure~\ref{1}).}
\label{2}
\end{center}
\end{figure}
Figure~\ref{1} illustrates the behavior of each of the contributions to $\langle d H_A/d\eta \rangle$ as functions of the Rindler coordinates $\xi_1$ and $\xi_2$ and with isotropic polarizations $\mu^{1} = \mu^{2} = \mu^{3}$. As discussed, each fixed $\xi$ represents the world line of an uniformly accelerated observer; lines of large positive $\xi$ are associated with weakly accelerated observers, while the hyperbolae that have large negative $\xi$ have a high proper acceleration. In this context, we see from figure~\ref{1} that for large negative $\xi$'s the contribution of vacuum fluctuations and radiation reaction present significantly different behaviors. In this limit the former acquires large positive values whereas the latter does not change appreciably, within the considered values in the plot, as compared with vacuum fluctuations. This implies that the generation of entanglement between highly accelerated atoms is much more sensitive to vacuum fluctuations as compared with the radiation-reaction effect. For atoms with lower proper accelerations, both effects approximately yield contributions of the same order.

For clarity one may consider the situation in which both atoms travel along the same hyperbolic trajectory $\xi_1 = \xi_2 = \xi$. In this case, they have the same proper acceleration $\alpha_1 = \alpha_2 = \alpha$. In this limit $\cosh\psi_{cd} \to 1$, see Eq.~(\ref{psi}). Expressions~(\ref{dd1}) and~(\ref{dd23}) reduce to:
\beq
\text{csch}^2\left(\psi_{12}\right)\,{\cal \widetilde{U}}_{k,12}(\omega,\alpha, \eta) \to \frac{\omega  \left(\alpha^2+\omega^2\right)}{3 \alpha^3}\,\mu_k
\eeq
with $k = 1, 2, 3$, $\text{csch}^2\left(\psi_{12}\right)\,{\cal \widetilde{U}}_{k,21} = \text{csch}^2\left(\psi_{12}\right)\,{\cal \widetilde{U}}_{k,12}$ in this limit and also ${\cal \widetilde{V}}_{i,12}(\omega, \alpha_1) = {\cal \widetilde{V}}_{i,12}(\omega, \alpha_2)$. This situation allows one to easily express the rate in terms of the proper time of the atoms by employing relation~(\ref{propt}). Hence the expressions of vacuum fluctuations and radiation reaction simplify to, respectively
\beq
\Biggl\langle \frac{d H_A}{d\tau} \Biggr\rangle_{VF} = \frac{\omega_0^4\,\boldsymbol\mu^2}{3\pi}
\left(\frac{\alpha^2}{\omega_0^2}+1\right)\,
\left(1+\frac{2}{e^{2\pi \omega_0/\alpha} - 1}\right),
\eeq
and
\beq
\Biggl\langle \frac{d H_A}{d\tau} \Biggr\rangle_{RR} = -\frac{\omega_0^4\,\boldsymbol\mu^2}{3\pi}
\left(\frac{\alpha^2}{\omega_0^2}+1\right).
\eeq
Let us investigate the behavior of each of the contributions to $\langle d H_A/d\tau \rangle$ as functions of the parameter $a$. This is depicted in figure~\ref{2}. We see from such a plot that for high values of $a$ (i.e., high proper accelerations) the differences between vacuum-fluctuation and radiation-reaction contributions become increasingly clear. The consequence of these considerations is that in principle one could plainly distinguish the contributions of vacuum fluctuations and radiation reaction for the formation of entangled atoms only for highly accelerated atoms. Incidentally we call attention for the fact that for the transition $|gg\rangle \to |\Psi^{-}\rangle$, the case with equal proper accelerations offers an interesting result: for both contributions of vacuum fluctuations and radiation reaction the cross correlations cancel the terms associated with isolated atoms. Hence $\langle d H_A/d\tau \rangle = 0$ for this transition. The Bell state $|\Psi^{-}\rangle$ cannot be created from ground-state atoms with the same proper acceleration. We remark that for the case of the transitions $|\Psi^{\pm}\rangle \to |gg\rangle$ and $|\Psi^{\pm}\rangle \to |ee\rangle$, i.e. degradation of entanglement, the results are qualitatively the same as those discussed above, including the fact that $\langle d H_A/d\tau \rangle = 0$ for the antisymmetric state in the situation of equal accelerations: the Bell state $|\Psi^{-}\rangle$ is stable with respect to radiative processes. Thus we recover a very known result in a complete different scenario, namely that atoms confined into a region much smaller than the optical wavelength, the antisymmetric Bell state $|\Psi^{-}\rangle$ can be regarded as a decoherence-free state~\cite{rep}. Furthermore, the results presented here should be compared with the outcomes of Ref.~\cite{china5}, even though they are working with a scalar field. In this case in which both atoms have the same proper acceleration we expect to find similar qualitative conclusions as those of such a reference.

\section{Conclusions and Perspectives}
\label{conclude}

There have been many works investigating quantum information theory within the framework of quantum field theory and quantum field theory in curved space-time. Several of such studies were implemented in a framework of open quantum systems. Following the discussion by Dalibard, Dupont-Roc, and Cohen-Tannoudji, we proposed to analyze the distinct contributions of vacuum fluctuations and radiation reaction to the quantum entanglement between two identical atoms in uniformly accelerated motion. We assume that both atoms are coupled to a quantum electromagnetic field. We have shown that the contributions of vacuum fluctuations and radiation reaction to the rate of variation of atomic anergy substantially differ for highly accelerated atoms. In addition, the creation of entanglement is possible for atoms initially prepared in the ground state. It is possible to envisage our results as the situation of two atoms coupled individually to two spatially separated cavities at different temperatures. For equal accelerations, we have found that the antisymmetric Bell state $|\Psi^{-}\rangle$ acts as a decoherence-free state

A natural extension of this paper is to consider the same setup studied here in the presence of boundaries. Since the presence of boundaries affect the vacuum fluctuations of the quantum field one should expect that the transition rates of atoms are modified in this situation. Hence, for entangled states, which have a non-local nature, it is interesting to ask how the transition rates are modified in a situation where translational invariance is broken. In turn, a related scenario is the investigation of uniformly accelerated qubits interacting with a massless scalar field via monopole interaction. Similar to the case studied here, we expect the radiation-reaction terms to possess some acceleration dependence as well as the appearance of non thermal contributions (which also depend on the acceleration) to the vacuum fluctuations of the field. However, we suspect that the emergence of such terms are connected solely with the presence of cross correlations between the atoms mediated by the scalar field, in contrast with the situation with the electromagnetic field. Another possible direction is to consider a black-hole space-time. One may argue that, close to event horizon the Schwarzschild metric takes the form of the Rindler line element. Therefore, in a certain sense the results discussed here can be extended to Schwarzschild black holes assuming that the atoms are close to the event horizon. In spite of this, the topic is more subtle than it would appear at first. Although entanglement is degraded near a Schwarzschild black hole as predicted in~\cite{fue}, there are important discrepancies arising due to the difference between an event horizon in Schwarzschild space-time and an acceleration horizon in Rindler space-time. Such subjects are under investigation and results will be published elsewhere.

\section*{acknowlegements}

This paper was supported by Conselho Nacional de Desenvolvimento Cientifico e Tecnol{\'o}gico do Brasil (CNPq).\\

\appendix

\section{Correlation functions of the electromagnetic fields in the vacuum}
\label{A}

In this appendix we shall briefly discuss free electromagnetic correlation functions. We consider a four dimensional unbounded Minkowski space-time. In addition, as previously mentioned we will work in the Coulomb gauge. Here we will not discuss the quantization of the electromagnetic field in detail. For a thorough analysis of quantum electrodynamics in the Coulomb gauge, we refer the reader the reference~\cite{cohen}. In turn, details concerning the correlations of the electromagnetic field for the case of
uniformly accelerated observers can be found in the comprehensive account by Takagi~\cite{takagi} and related references
therein. Instead, we limit ourselves to briefly discuss the evaluation of all relevant electromagnetic correlation functions in the vacuum employed in the text.

We start our discussion by presenting the Hadamard's elementary function associated with the electric field. It is given by
\begin{widetext}
\bea
D^{ij}(x,x') &=& \langle 0 |\{E^{i}(x), E^{j}(x')\}| 0\rangle = \left(\frac{\partial}{\partial t}\frac{\partial}{\partial t'}\delta^{ij} - \frac{\partial}{\partial x_i}\frac{\partial}{\partial x_j'}\right)\,D^{(1)}(t-t',{\bf x} - {\bf x}'),
\nn\\
&=& -\left(\frac{\partial^2}{\partial \eta^2}\delta^{ij} - \frac{\partial}{\partial \rho_i}\frac{\partial}{\partial \rho_j}\right)\,D^{(1)}(\eta,\mbox{\small\mathversion{bold}${\rho}$}),
\eea
where
\beq
D^{(1)}(t - t',{\bf x} - {\bf x}') = -\frac{1}{4\pi^2}\left\{\frac{1}{\left[(t - t' - i\epsilon)^2 - |{\bf x} - {\bf x}'|^2\right]} + \frac{1}{\left[(t - t' + i\epsilon)^2 - |{\bf x} - {\bf x}'|^2\right]}\right\}.
\eeq
After calculating the derivatives one gets
\bea
D^{ij}(x,x') &=& -\frac{1}{\pi^2}\biggl\{\frac{2(\Delta{\bf x})^i(\Delta{\bf x})^j - \delta^{ij}\left[|\Delta{\bf x}|^2 + (\Delta t - i\epsilon)^2\right]}{\left[(t - t' - i\epsilon)^2 - |{\bf x} - {\bf x}'|^2\right]^3}
\nn\\
&+& \frac{2(\Delta{\bf x})^i(\Delta{\bf x})^j - \delta^{ij}\left[|\Delta{\bf x}|^2 + (\Delta t+i\epsilon)^2\right]}{\left[(t - t' + i\epsilon)^2 - |{\bf x} - {\bf x}'|^2\right]^3}\biggr\}.
\label{hada}
\eea
The Pauli-Jordan function of the electric field is given by
\bea
\Delta^{ij}(x,x') &=& \langle 0 |[E^{i}(x), E^{j}(x')]| 0\rangle = -i\left(\frac{\partial}{\partial t}\frac{\partial}{\partial t'}\delta^{ij} - \frac{\partial}{\partial x_i}\frac{\partial}{\partial x_j'}\right)\,D(t - t',{\bf x} - {\bf x}')
\nn\\
&=&i\left(\frac{\partial^2}{\partial \zeta^2}\delta^{ij} - \frac{\partial}{\partial \rho_i}\frac{\partial}{\partial \rho_j}\right)\,D(\eta,\mbox{\small\mathversion{bold}${\rho}$}),
\eea
where $\zeta = t - t'$, $\mbox{\small\mathversion{bold}${\rho}$} = {\bf x} - {\bf x}'$ and
\bea
D(t,{\bf x}) &=& \frac{1}{2\pi}\,\sgn (t)\delta(|{\bf x}|^2 - t^2) = \frac{1}{4\pi|{\bf x}|}\Bigl[\delta(|{\bf x}| - t) - \delta(|{\bf x}| + t)\Bigr]
\nn\\
&=&\, \frac{1}{(2\pi)^2|{\bf x}|}\lim_{\epsilon \to 0}\left[\frac{\epsilon}{\epsilon^2+(|{\bf x}| - t)^2} - \frac{\epsilon}{\epsilon^2+(|{\bf x}| + t)^2}\right].
\label{pj}
\eea
A detailed analysis of the function $D(t,{\bf x})$ can be found in reference~\cite{cohen}. By evaluating the derivatives one obtains an explicit form for the Pauli-Jordan function in terms of the delta functions and their derivatives:	
\bea
\Delta^{ij}(x,x') &=& \frac{i}{4\pi}\Biggl\{\left(\frac{3(\Delta{\bf x})^i(\Delta{\bf x})^j}{|\Delta{\bf x}|^2} - \delta^{ij}\right)\Biggl[\frac{\delta'(|\Delta{\bf x}| - \Delta t) - \delta'(|\Delta{\bf x}| + \Delta t)}{|\Delta{\bf x}|^2}
\nn\\
&-& \left(\frac{\delta(|\Delta{\bf x}| - \Delta t) - \delta(|\Delta{\bf x}| + \Delta t)}{|\Delta{\bf x}|^3}\right)\Biggr]
\nn\\
&-&\left(\frac{(\Delta{\bf x})^i(\Delta{\bf x})^j}{|\Delta{\bf x}|^2} - \delta^{ij}\right)\left[\frac{\delta''(|\Delta{\bf x}| - \Delta t) - \delta''(|\Delta{\bf x}| + \Delta t)}{|\Delta{\bf x}|}\right]\Biggr\},
\label{pauli}
\eea
where $\Delta{\bf x} = \mbox{\small\mathversion{bold}${\rho}$}$ and $\Delta t = \zeta$. The above derivatives of the delta function are to be understood as a distributional derivative. One can also express the Pauli-Jordan function based on the derivation of the Hadamard's elementary function:
\bea
\Delta^{ij}(x,x') &=& -\frac{1}{\pi^2}\biggl\{\frac{2(\Delta{\bf x})^i(\Delta{\bf x})^j - \delta^{ij}\left[|\Delta{\bf x}|^2 + (\Delta t - i\epsilon)^2\right]}{\left[(t - t' - i\epsilon)^2 - |{\bf x} - {\bf x}'|^2\right]^3}
\nn\\
&-& \frac{2(\Delta{\bf x})^i(\Delta{\bf x})^j - \delta^{ij}\left[|\Delta{\bf x}|^2 + (\Delta t+i\epsilon)^2\right]}{\left[(t - t' + i\epsilon)^2 - |{\bf x} - {\bf x}'|^2\right]^3}\biggr\}.
\label{pauli-e}
\eea
Now let us present the commutators associated with the magnetic field. From standard calculations, it is easy to show that $\langle 0 |B^{i}(x) B^{j}(x')| 0\rangle = \langle 0 |E^{i}(x) E^{j}(x')| 0\rangle$ and therefore
\beq
\langle 0 |[B^{i}(x), B^{j}(x')]| 0\rangle = \langle 0 |[E^{i}(x), E^{j}(x')]| 0\rangle,\langle 0 |\{B^{i}(x), B^{j}(x')\}| 0\rangle = \langle 0 |\{E^{i}(x), E^{j}(x')\}| 0\rangle.
\eeq
On the other hand, the correlation function of both fields is given by
\beq
\langle 0 |E^{i}(x) B^{j}(x')| 0\rangle = -i\sum_{m}\epsilon^{ijm}\frac{\partial}{\partial \eta} \frac{\partial}{\partial \rho_{m}} D^{+}(\eta,\mbox{\small\mathversion{bold}${\rho}$}),
\eeq
where $\epsilon^{ijm}$ is the Levi-Civita tensor density and
\beq
D^{+}(x,x') = \frac{i}{(2\pi)^3}\int \frac{d^3k}{2\omega_{{\bf k}}}e^{i [{\bf k}\cdot ({\bf x}-{\bf x}')-\omega_{{\bf k}}(t-t')]}
= -\frac{i}{4\pi^2}\frac{1}{\left[(t - t' - i\epsilon)^2 - |{\bf x} - {\bf x}'|^2\right]}.
\eeq
Therefore
\beq
\langle 0 |E^{i}(x) B^{j}(x')| 0\rangle = \frac{2}{\pi^2}\sum_{m}\epsilon^{ijm} \frac{(\Delta{\bf x})_m(\Delta t - i\epsilon)}{\left[(t - t' - i\epsilon)^2 - |{\bf x} - {\bf x}'|^2\right]^3}.
\label{eb}
\eeq
Hence, the symmetric correlation function is given by
\beq
\langle 0 |\{E^{i}(x), B^{j}(x')\}| 0\rangle = \frac{2}{\pi^2}\sum_{m}\epsilon^{ijm} \left\{\frac{(\Delta{\bf x})_m(\Delta t - i\epsilon)}{\left[(t - t' - i\epsilon)^2 - |{\bf x} - {\bf x}'|^2\right]^3} + \frac{(\Delta{\bf x})_m(\Delta t + i\epsilon)}{\left[(t - t' + i\epsilon)^2 - |{\bf x} - {\bf x}'|^2\right]^3}\right\},
\label{eb-sym}
\eeq
and the commutator reads
\bea
\langle 0 |[E^{i}(x), B^{j}(x')]| 0\rangle &=& \frac{2}{\pi^2}\sum_{m}\epsilon^{ijm} \left\{\frac{(\Delta{\bf x})_m(\Delta t - i\epsilon)}{\left[(t - t' - i\epsilon)^2 - |{\bf x} - {\bf x}'|^2\right]^3} - \frac{(\Delta{\bf x})_m(\Delta t + i\epsilon)}{\left[(t - t' + i\epsilon)^2 - |{\bf x} - {\bf x}'|^2\right]^3}\right\}
\nn\\
&=&\frac{i}{4\pi}\sum_{m}\epsilon^{ijm}\frac{(\Delta{\bf x})_m}{|\Delta{\bf x}|}\Biggl[\frac{\delta''(|\Delta{\bf x}| - \Delta t) + \delta''(|\Delta{\bf x}| + \Delta t)}{|\Delta{\bf x}|}
\nn\\
&-&\, \left(\frac{\delta'(|\Delta{\bf x}| - \Delta t) + \delta'(|\Delta{\bf x}| + \Delta t)}{|\Delta{\bf x}|^2}\right)\Biggr],
\label{eb-comm}
\eea
where the last line follows from the limit $\epsilon \to 0$.

The statistical functions of the fields have to be evaluated along the trajectory of the atoms. From the results above one may evaluate the Hadamard's elementary functions found in equation~(\ref{vfha3}). For the case in which the atoms are uniformly accelerated, the discussion is involved. As a basis of our analysis, we consider the following system of reference frames: the inertial laboratory frame $S$ with Minkowski coordinates $(t,{\bf x})$, a set of instantaneous inertial frames $S_{\tau}^{1}$ defined at each value of the proper time of atom 1 and another set of instantaneous inertial frames $S_{\tau}^{2}$ defined at each value of the proper time of atom 2. The latter employ Rindler coordinates $(\eta, \xi,{\bf x}_{\perp})$. In this way, the transformations from the inertial frame $S$ to one of the frames $S_{\tau}^{i}$, $i=1,2$, for an arbitrary tensor field $T^{\mu\cdots\sigma}_{\,\,\,\nu\cdots\rho}$ are given by standard Lorentz transformations:
\beq
T^{\mu\cdots\sigma}_{\,\,\,\nu\cdots\rho}(\tau) = \Lambda^{\mu}_{\,\,\,\alpha}(\tau)\cdots\Lambda^{\sigma}_{\,\,\,\beta}(\tau)\,\Lambda_{\nu}^{\,\,\,\delta}(\tau)\cdots\Lambda_{\rho}^{\,\,\,\gamma}(\tau)\,T^{\alpha\cdots\beta}_{\,\,\,\delta\cdots\gamma}(x(\tau)),
\label{tensor}
\eeq
where
\beq
\Lambda = \left( \begin{array}{cccc}
\gamma(\tau) & -\beta(\tau)\gamma(\tau) & 0 & 0 \\
-\beta(\tau)\gamma(\tau) & \gamma(\tau) & 0 & 0 \\
0 & 0 & 1 & 0\\
0 & 0 & 0 & 1
\end{array} \right),
\label{lambda}
\eeq
where $\beta(\tau(\eta)) = (dx_1/d\eta)/(dt/d\eta) = \tanh a\eta$ and $\gamma(\tau(\eta)) = 1/(1- \beta^2)^{1/2} = \cosh a\eta$. Remember that $\eta$ and $\tau$ are related by expression~(\ref{propt}). For instance, for the electromagnetic field strength tensor $F_{\mu\nu} = \partial_{\mu}A_{\nu} - \partial_{\nu}A_{\mu}$ one has
\beq
F^{\alpha\beta}(\tau) = F^{\mu\nu}(x(\tau))\Lambda^{\alpha}_{\,\,\,\mu}(\tau)\Lambda^{\beta}_{\,\,\,\nu}(\tau),
\label{tensor1}
\eeq
therefore one can obtain the associated transformation for the correlation function of the electromagnetic field strength tensor as follows
\beq
\langle 0 |F^{\alpha\beta}(\tau) F^{\gamma\delta}(\tau')| 0\rangle = \langle 0 |F^{\mu\nu}(x(\tau)) F^{\rho\sigma}(x(\tau'))| 0\rangle\Lambda^{\alpha}_{\,\,\,\mu}(\tau)\Lambda^{\beta}_{\,\,\,\nu}(\tau)\Lambda^{\gamma}_{\,\,\,\rho}(\tau')\Lambda^{\delta}_{\,\,\,\sigma}(\tau').
\label{tensor2}
\eeq
From the usual relations $E^{i} = F^{i0}$, and $B_{i} = -\epsilon_{ijk}F^{jk}/2$,
and the results derived above one can obtain the correlation functions for the electromagnetic fields ${\bf E}$ and ${\bf B}$. Then it is straightforward to obtain the associated transformations for the Hadamard's elementary functions and also for the Pauli-Jordan functions. For simplicity, we consider the case in which both atoms have the same spatial coordinates which are orthogonal to the direction of the acceleration. Since $x = x(\eta)$ and $x' = x(\eta')$ from equation~(\ref{hada}) and employing equation~(\ref{acc}) as well as the above coordinate transformations one has the following explicit form for the Hadamard's elementary function ($a e^{-a\xi_1} = \alpha_1$,$a e^{-a\xi_2} = \alpha_2$) appearing in equation~(\ref{vfha3}):
\bea
D_{ij}[x_c(\eta),x_d(\eta')] &=& \frac{a^4 e^{-2 a(\xi_c + \xi_d)}}{4\pi^2}\Biggl\{\frac{\delta_{ij} \nu_{i,cd}(a(\eta-\eta'),\epsilon)}{\left[\cosh (a(\eta-\eta')- ia\epsilon)- \cosh\psi_{cd} \right]^3}
+\frac{\delta_{ij} \nu_{i,cd}^{*}(a(\eta-\eta'),\epsilon)}{\left[\cosh (a(\eta-\eta')+ ia \epsilon)- \cosh\psi_{cd} \right]^3}\Biggr\},
\nn\\
\label{hada-acc2}
\eea
where there is no implicit summation over repeated indices and we have defined
\bea
\nu_{1,cd}(a(\eta-\eta'),\epsilon) &=& \cosh (a(\eta-\eta') - ia\epsilon)- \cosh\psi_{cd}
\nn\\
\nu_{2,cd}(a(\eta-\eta'),\epsilon) &=& \nu_{3,cd}^{(\pm)} = \cosh (a(\eta-\eta') - ia\epsilon)\cosh\psi_{cd} - 1
\eea
with $\psi_{cd} = \psi({\bf x}_{c,\perp}, \xi_c,{\bf x}_{d,\perp}, \xi_d)$ defined through the following equation:
\beq
\cosh\psi({\bf x}_{c,\perp}, \xi_c,{\bf x}_{d,\perp}, \xi_d) = 1 + \frac{(e^{a\xi_c} - e^{a\xi_d})^2}{2e^{a(\xi_c + \xi_d)}}.
\label{psi}
\eeq
In order to derive such expressions use has been made of known hyperbolic identities~\cite{abram}. In addition, we have absorbed a positive function of $\eta, \eta'$ as well as a positive function of $\xi_1, \xi_2$ into $\epsilon$. Incidentally, we mention that in order to reach the above Wightman functions we have redefined the regulator $\epsilon$. For a finite time interaction between the atoms and the field, this leads to a non-trivial change in the interpretation of $\epsilon$. For more details on this subject, see Ref.~\cite{LinHu2010}. Available discussion of the renormalization of the epsilon-regularization and alternatives
to it can also be found in Refs.~\cite{langlois:05,mar2}.

On the other hand, from the above transformations one can compute the Pauli-Jordan function for this case which are present in equation~(\ref{rrha3}):
\bea
\Delta_{ij}[x_c(\eta),x_d(\eta')] &=& \frac{a^4 e^{-2 a(\xi_c + \xi_d)}}{4\pi^2}\Biggl\{\frac{\delta_{ij} \nu_{i,cd}(a(\eta-\eta'),\epsilon)}{\left[\cosh (a(\eta-\eta')- ia\epsilon)- \cosh\psi_{cd} \right]^3}
-\frac{\delta_{ij} \nu_{i,cd}^{*}(a(\eta-\eta'),\epsilon)}{\left[\cosh (a(\eta-\eta')+ ia\epsilon)- \cosh\psi_{cd} \right]^3}\Biggr\}.
\nn\\
\label{pauli-acc2}
\eea
%

\section{Evaluation of the contributions of vacuum fluctuations and radiation reaction to $\langle d H_A/d\eta \rangle$}
\label{B}

\begin{figure}
\centering\includegraphics[width=1.0\linewidth]{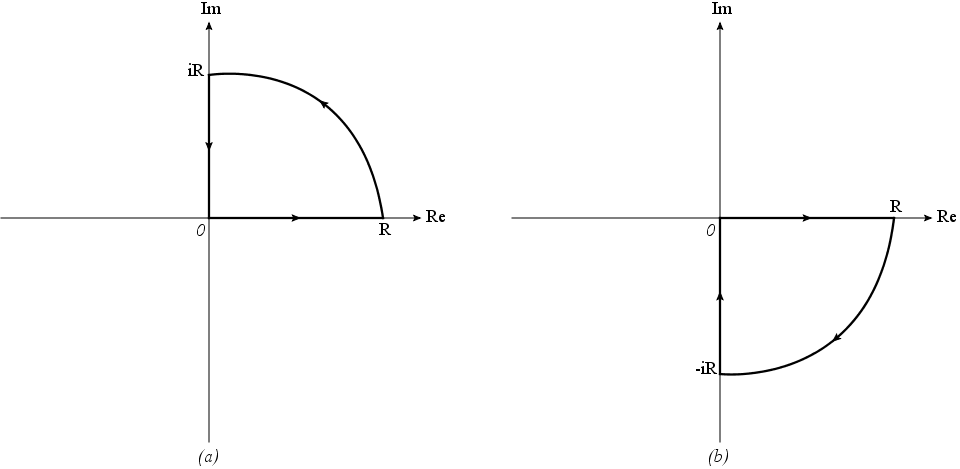}
        \caption{Contours for the evaluation of the integrals of equations~(\ref{energy-flu}) and~(\ref{energy-rad}) ($R \to \infty$). Exceptionally, this figure was developed with the JaxoDraw Java program.}
\label{3}
\end{figure}

In this Appendix we calculate in detail the contributions of vacuum fluctuations and radiation reaction to the rate of variation of energy of the two-atom system. From the results derived in the Appendix~\ref{A}, one may compute all the relevant correlation functions of the electric field which appears in equations~(\ref{vfha3}) and~(\ref{rrha3}). Namely, such contributions to spontaneous emission are computed by inserting in such expressions the statistical functions of the two atom system, given by equations~(\ref{susa1}) and~(\ref{cora1}), and the electromagnetic-field statistical functions given by~(\ref{hada-acc2}) and~(\ref{pauli-acc2}). Initially let us present the contributions coming from the vacuum fluctuations. Performing a simple change of variable $u = a(\eta - \eta')$, these can be expressed as, with $\Delta\eta = \eta - \eta_0$:
\bea
\Biggl\langle \frac{d H_A}{d\eta} \Biggr\rangle_{VF} &=& -\frac{a^3}{8\pi^2}\sum_{\omega'}\sum_{c,d = 1}^{2} \Delta\omega\,e^{- a(\xi_c + \xi_d)}\,\exp\left[\frac{i\Delta\omega(\alpha_d -\alpha_c)a\eta}{\alpha_c\alpha_d}\right]
\nn\\
&\times&\int_{0}^{a\Delta\eta}du \left(e^{i\Delta\omega u/\alpha_d}
+ e^{-i\Delta\omega u/\alpha_c}\right)\Biggl[\frac{\delta_{ij} T^{ij}_{cd}(\omega,\omega'; u,\epsilon)}{\left(\cosh (u- i\epsilon)- \cosh\psi_{cd} \right)^3}
+\frac{\delta_{ij} T^{ij\,*}_{cd}(\omega,\omega'; u,\epsilon)}{\left(\cosh (u+ i\epsilon)- \cosh\psi_{cd} \right)^3}\Biggr],
\label{energy-flu}
\eea
where $a$ was absorbed into $\epsilon$ and we have defined
\beq
T^{ij}_{cd}(\omega,\omega'; u,\epsilon) = {\cal A}^{ij}_{cd}(\omega,\omega') \nu_{i,cd}(u,\epsilon),
\label{b2}
\eeq
with no implicit summation on repeated indices in equation~(\ref{b2}). The quantities $\nu_{i,cd}$ and $\psi_{cd} = \psi({\bf x}_{c,\perp}, \xi_c,{\bf x}_{d,\perp}, \xi_d)$ are defined in Appendix~\ref{A} whereas the generalized atomic transition dipole moment ${\cal A}^{ij}_{cd}(\omega,\omega')$ is defined in equation~(\ref{aa}). All integrals can be evaluated within the techniques of contour integration by noting that ($z > 0$)
$$
\int_{0}^{\infty}\,dr\,[\cdots]\, = \int_{0}^{z}\,dr\,[\cdots] + \int_{z}^{\infty}\,dr\,[\cdots].
$$
The strategy goes as follows. Consider the integrals:
$$
I_1 = \int_{0}^{\infty}du\frac{e^{i\Delta\omega u/\alpha_d}\,\nu_{i,cd}(u,\epsilon)}{\sinh^3\left(\frac{u - i\epsilon + \psi_{cd}}{2} \right)\sinh^3\left(\frac{u -  i\epsilon - \psi_{cd}}{2} \right)},
$$
and
$$
I_2 = \int_{0}^{\infty}du\frac{e^{-i\Delta\omega u/\alpha_c}\,\nu_{i,cd}(u,\epsilon)}{\sinh^3\left(\frac{u - i\epsilon + \psi_{cd}}{2} \right)\sinh^3\left(\frac{u -  i\epsilon - \psi_{cd}}{2} \right)}.
$$
Let us take a look at $I_1$. We use the contours in the complex $u$-plane illustrated in figure~\ref{3}. For $\Delta\omega > 0$, the poles located inside the contour considered are given by $u_n = i\epsilon + \psi_{cd} + 2\pi i n$, where $n \geq 0$ is an integer and $\psi_{cd} > 0$. The associated contour integral can be splitted up as
$$
\oint_{\Delta\omega > 0}du\,[\cdots] = \int_{0}^{R}du\,[\cdots] + \int_{c_R}du\,[\cdots] + \int_{iR}^{0}du\,[\cdots] = 2\pi i\, \sum_{n = 0}^{\infty}\textrm{Res}(u_n),
$$
where $c_R$ is the quarter of the circle depicted in plot $(a)$ of Fig.~\ref{3}. On the other hand, for $\Delta\omega < 0$ the poles inside the contour are given by $u_n = \psi_{cd} - 2\pi i n$, where $n > 0$. Similarly, such a contour can be splitted up as
$$
\oint_{\Delta\omega < 0}du\,[\cdots] = \int_{0}^{R}du\,[\cdots] + \int_{c_R'}du\,[\cdots] + \int_{-iR}^{0}du\,[\cdots] = -2\pi i\,\sum_{n = 1}^{\infty}\textrm{Res}(u_n),
$$
where $c_R'$ is the quarter of the circle depicted in plot $(b)$ of Fig.~\ref{3}. The integrals along $c_R$ and $c_R'$ vanish for $R \to \infty$ in virtue of the Jordan's lemma. Similar considerations can be settled to $I_2$. It is clear that same arguments apply for the integral associated with $\nu_{i,cd}^{*}(u,\epsilon)$. Hence, after some algebra one gets, for $\Delta\eta > \psi_{cd}/a > 0$ and $\epsilon \to 0$
\bea
&& \Biggl\langle \frac{d H_A}{d\eta} \Biggr\rangle_{VF} =  -\frac{a^3}{4\pi}\sum_{\omega'}\sum_{c,d = 1}^{2}e^{- a(\xi_c + \xi_d)}\,\exp\left[\frac{i\Delta\omega(\alpha_d -\alpha_c)a\eta}{\alpha_c\alpha_d}\right]\,\text{csch}^2\left(\psi_{cd}\right)
\nn\\
&\times&\,\Biggl\{\Delta\omega\,{\cal J}^{\omega\omega'}_{cd}(\Delta\omega,\Delta\eta)
+\theta(\Delta\omega)\Delta\omega\Biggl[\delta_{ij}\biggl(U^{ij}_{cd}(\Delta\omega,\alpha_d)
+ V^{ij}_{cd}(\Delta\omega,\alpha_d)
+\frac{2\,U^{ij}_{cd}(\Delta\omega,\alpha_d) + V^{ij}_{cd}(\Delta\omega,\alpha_d)}
{e^{2\pi \Delta\omega /\alpha_d} - 1}\biggr)
\nn\\
&+&\,\delta_{ij}\biggl(\Bigl(U^{ij}_{cd}(\Delta\omega,\alpha_c)\Bigr)^{*} + \Bigl(V^{ij}_{cd}(\Delta\omega,\alpha_c)\Bigr)^{*}
+\frac{2\,\Bigl(U^{ij}_{cd}(\Delta\omega,\alpha_c)\Bigr)^{*} + \Bigl(V^{ij}_{cd}(\Delta\omega,\alpha_c)\Bigr)^{*}}
{e^{2\pi \Delta\omega /\alpha_c} - 1}\biggr)\Biggr]
\nn\\
&-&\, \theta(-\Delta\omega)|\Delta\omega|\Biggl[\delta_{ij}\biggl(\Bigl(U^{ij}_{cd}(|\Delta\omega|,\alpha_d)\Bigr)^{*}
+ \Bigl(V^{ij}_{cd}(|\Delta\omega|,\alpha_d)\Bigr)^{*}
+\frac{2\,\Bigl(U^{ij}_{cd}(|\Delta\omega|,\alpha_d)\Bigr)^{*}
+ \Bigl(V^{ij}_{cd}(|\Delta\omega|,\alpha_d)\Bigr)^{*}}{e^{2\pi |\Delta\omega| /\alpha_d} - 1}\biggr)
\nn\\
&+&\,\delta_{ij}\biggl(U^{ij}_{cd}(|\Delta\omega|,\alpha_c) + V^{ij}_{cd}(|\Delta\omega|,\alpha_c)
+\frac{2\,U^{ij}_{cd}(|\Delta\omega|,\alpha_c) + V^{ij}_{cd}(|\Delta\omega|,\alpha_c)}
{e^{2\pi |\Delta\omega| /\alpha_c} - 1}\biggr)\Biggr]\Biggr\},
\label{vacuum}
\eea
where we have defined
\beq
{\cal J}^{\omega\omega'}_{cd}(\Delta\omega, \Delta\eta) = - \frac{1}{8\pi\text{csch}^2\psi_{cd}}\int_{a\Delta\eta}^{\infty}du \,\left(e^{i\Delta\omega u/\alpha_d}
+ e^{-i\Delta\omega u/\alpha_c}\right)\frac{\delta_{ij} T^{ij}_{cd}(\omega,\omega'; u, 0)}{\sinh^3\left(\frac{u + \psi_{cd}}{2} \right)\sinh^3\left(\frac{u - \psi_{cd}}{2} \right)}
\eeq
and also
\beq
U^{ij}_{cd}(\Delta\omega,\alpha) = {\cal A}^{ij}_{cd}(\omega,\omega')\,{\cal U}_{i,cd}(\Delta\omega,\alpha)\,
\label{def-u}
\eeq
\beq
V^{ij}_{cd}(\Delta\omega, \alpha) =  \frac{i}{\pi\text{csch}^2\psi_{cd}}\int_{0}^{2\pi}du \,e^{-\Delta\omega u/\alpha}\frac{T^{ij}_{cd}(\omega,\omega'; iu, 0)}{\left(\cos u - \cosh\psi_{cd} \right)^3},
\label{def-v}
\eeq
with no implicit summation on repeated indices in equation~(\ref{def-u}), and
\beq
{\cal U}_{1,cd}(\Delta\omega,\alpha) = -\exp\left(\frac{i\Delta\omega \psi_{cd}}{\alpha}\right)\,\left[\frac{\Delta \omega}{\alpha}+ i\,\coth\psi_{cd}\right]
\label{d1}
\eeq
\beq
{\cal U}_{2,cd}(\Delta\omega,\alpha) = {\cal U}_{3,cd}(\Delta\omega,\alpha)
=  \frac{1}{2}\,\exp\left(\frac{i\Delta\omega \psi_{cd}}{\alpha}\right)\,\left[\frac{\Delta\omega}{\alpha}\cosh \psi_{cd}
+ i\,\text{csch}\psi_{cd}\left(1- \left(\frac{\Delta\omega}{\alpha}\right)^2 \sinh^2(\psi_{cd})\right) \right].
\label{d23}
\eeq
In the above equations we used the fact that $\nu_{i,cd}^{*}(u,0) = \nu_{i,cd}(u,0)$. Now let us evaluate the radiation-reaction contributions. As above we perform a simple change of variable $u = a(\eta - \eta')$. One gets
\bea
 \Biggl\langle \frac{d H_A}{d\eta} \Biggr\rangle_{RR} && = -\frac{a^3}{8\pi^2}\sum_{\omega'}\sum_{c,d = 1}^{2}e^{- a(\xi_c + \xi_d)}\Delta\omega
\nn\\
&&\times\int_{0}^{a\Delta\eta}du \left(e^{i\Delta\omega u/\alpha_d}
- e^{-i\Delta\omega u/\alpha_c}\right)\Biggl[\frac{\delta_{ij} T^{ij}_{cd}(u,\epsilon,\eta)}{\left(\cosh (u- i\epsilon)- \cosh\psi_{cd} \right)^3}
-\frac{\delta_{ij} T^{ij\,*}_{cd}(u,\epsilon,\eta)}{\left(\cosh (u+ i\epsilon)- \cosh\psi_{cd} \right)^3}\Biggr].
\label{energy-rad}
\eea
Using the results derived above, one has, with $\Delta\eta > \psi_{cd}/a > 0$ and $\epsilon \to 0$:
\bea
\Biggl\langle \frac{d H_A}{d\eta} \Biggr\rangle_{RR} &=& -\frac{a^3}{4\pi}\sum_{\omega'}\sum_{c,d = 1}^{2}e^{- a(\xi_c + \xi_d)}\,\exp\left[\frac{i\Delta\omega(\alpha_d -\alpha_c)a\eta}{\alpha_c\alpha_d}\right]\,\text{csch}^2\left(\psi_{cd}\right)
\nn\\
&\times&\,\Biggl\{\theta(\Delta\omega)\Delta\omega\Biggl[\delta_{ij}U^{ij}_{cd}(\Delta\omega,\alpha_d)
+\delta_{ij}\Bigl(U^{ij}_{cd}(\Delta\omega,\alpha_c)\Bigr)^{*}\Biggr]
\nn\\
&+&\, \theta(-\Delta\omega)|\Delta\omega|\Biggl[\delta_{ij}\Bigl(U^{ij}_{cd}(|\Delta\omega|,\alpha_d)\Bigr)^{*}
+\delta_{ij}U^{ij}_{cd}(|\Delta\omega|,\alpha_c) \Biggr]\Biggr\}.
\label{radiation}
\eea
\end{widetext}

\end{document}